# Voltage-controlled inversion of tunnel magnetoresistance in epitaxial Nickel/Graphene/MgO/Cobalt junctions

F. Godel[1], M. Venkata Kamalakar[1,2], B. Doudin[1], Y. Henry[1], D. Halley[1]* and J.-F. Dayen[1]*.

1. Institut de Physique et Chimie des Matériaux de Strasbourg (IPCMS), UMR 7504 CNRS-UdS, 23 rue du Loess, 67034 Strasbourg, France.
2. Department of Microtechnology and Nanoscience, Chalmers University of Technology, SE-41296, Göteborg, Sweden.
* halley@ipcms.unistra.fr , dayen@ipcms.unistra.fr

**Abstract :** *We report on the fabrication and characterization of vertical spin-valve structures using a thick epitaxial MgO barrier as spacer layer and a graphene-passivated Ni film as bottom ferromagnetic electrode. The devices show robust and scalable tunnel magnetoresistance, with several changes of sign upon varying the applied bias voltage. These findings are explained by a model of phonon-assisted transport mechanisms that relies on the peculiarity of the band structure and spin density of states at the hybrid graphene/Ni interface.*

Spin-based memory and logic devices are the subject of an intense research activity motivated by the perspective to overcome power, performance and architectural bottlenecks of CMOS-based devices. Among potential material candidates in this field, graphene (Gr) carries great expectations because of its unique electronic transport properties. So far, graphene has been employed mainly in "lateral" spintronic devices, where ferromagnetic electrodes are deposited on top of graphene and electron current flows in the plane of the carbon sheet [1, 2, 3]. In such devices, oxide tunnel barriers (MgO or $Al_2O_3$) are often inserted between graphene and the ferromagnetic metals to overcome the conductance mismatch problem [4, 5], allowing spin-polarized electrons to be efficiently injected into or extracted from graphene. Most of the experimental work carried out so far has been aimed at elucidating the spin relaxation mechanisms in graphene, while the properties of the Gr|ferromagnet and Gr|oxide interfaces has remained essentially unexplored.

Mastering the spin filtering effects at interfaces is highly important for applications, as these are the cornerstones of many spintronic devices. In seminal first principle studies, Karpan *et al.* [6, 7] predicted that on increasing the number of carbon layers, large spin filtering efficiency should take place at the interface between few-layer graphene and ferromagnetic electrodes of (111) fcc or (0001) hcp nickel or cobalt. This was ascribed to the fact that the electronic structures of the two materials only overlap for the minority spin direction, in those parts of the reciprocal space corresponding to the *K* point of graphene, and that only minority electrons should therefore be transmitted from the metal surface into graphene. A first realization of graphene based current-perpendicular-to-plane (CPP) spintronic devices was recently reported [8]. Transferred single layer graphene was used as a tunnel barrier between Co and NiFe polycrystalline electrodes and positive tunnel magnetoresistance (TMR) values up to 2 % at 4 K were measured. Similar results were also reported for transferred graphene sandwiched between pairs of Co [9] and NiFe [10] electrodes.

Concomitantly, CPP spin-valve effects were demonstrated in devices containing few layer graphene grown directly onto nickel by chemical vapor deposition (CVD) [11]. These so-called Graphene Passivated Ferromagnetic Electrodes (GPFE) are particularly appealing candidates for spin-valve electrodes as they are intrinsically oxidation-resistant. A negative TMR was observed at 1.4 K for GPFE/$Al_2O_3$/Co stacks under bias voltages of +100 mV and -100 mV [11]. Assuming implicitly that Ni was (111)-oriented, the authors related the negative TMR to the theoretical prediction made by Karpan *et al.* [6, 7] of a negative tunneling spin polarization at the *K* point of the GPFE Brillouin zone. However, the voltage bias dependence of the TMR, which is critically important to understand the interfacial effects [12, 13, 14], has not been studied. Such bias dependence investigation is of primary importance in view of the recent first principle non-equilibrium transport calculations, which predict a non-trivial dependence of the spin polarization with varying applied voltage at Gr|Co(111) and Gr|Ni(111) interfaces [15].

In this paper, we present a detailed low temperature study of the electron transport properties of carefully characterized epitaxial (111)-oriented GPFE/MgO/Co tunnel junctions, with special emphasis on the voltage dependence of the tunnel magnetoresistance. Studies performed on several samples with surface area ranging from 1 µm$^2$ to 1000 µm$^2$ provide confidence that the reported observations are robust and scalable. On varying the bias voltage, the TMR ratio systematically shows three distinct regimes, along with a number of sign reversals. The transitions between regimes are interpreted as the opening/closing of spin-polarized conduction channels, among which phonon-assisted channels that allow electrons to overcome the constraint of tunneling with perpendicular-to-interface momentum imposed by the thick MgO barrier. The observed bias dependence of the TMR ratio is consistent with recent theoretical results of the band structure and the spin density of states at the hybrid Ni/Gr interface [6, 7].

The tunnel junctions were fabricated starting from commercial 1-7 layers graphene grown on 200 nm thick Ni film by CVD [16]. A 3 nm thick MgO tunnel barrier was then deposited at 100°C by molecular beam epitaxy (MBE) under ultrahigh vacuum (UHV), with pressure in the $10^{-8}$ mbar range [17]. To ensure low roughness and strong sticking of the oxide layer on graphene, 0.12 nm of Ti were dusted on the substrate prior to MgO deposition [18]. During MgO evaporation, the latter decomposes into atomic Mg and O species [19], resulting in a ~$5.10^{-8}$ torr oxygenated atmosphere that oxidizes Ti into $TiO_2$ [17, 20]. This small amount of titanium oxide improves the uniformity of the MgO layer [18]. Moreover titanium oxide based tunnel barriers have also been used in the past for efficient spin injection and detection [21, 22].

The MgO layer is, of course, of primary importance for spin transport but it also protects graphene from contamination during the subsequent patterning steps. Square holes were defined by electron lithography in a 150 nm layer of PMMA resist spin coated on MgO to define effective contact surface area ranging from 1 µm² to 1000 µm². The top ferromagnetic electrode, consisting of 50 nm of Co capped with 3 nm of Pd, was then deposited by UHV MBE through a shadow mask with slightly sub-millimeter size square apertures centered on the previously defined holes. Finally, Ti(10 nm)/Au(60 nm) top electrical contacts were formed by e-beam evaporation through the same mask. The final spin-valve design is depicted in Fig 1(a).

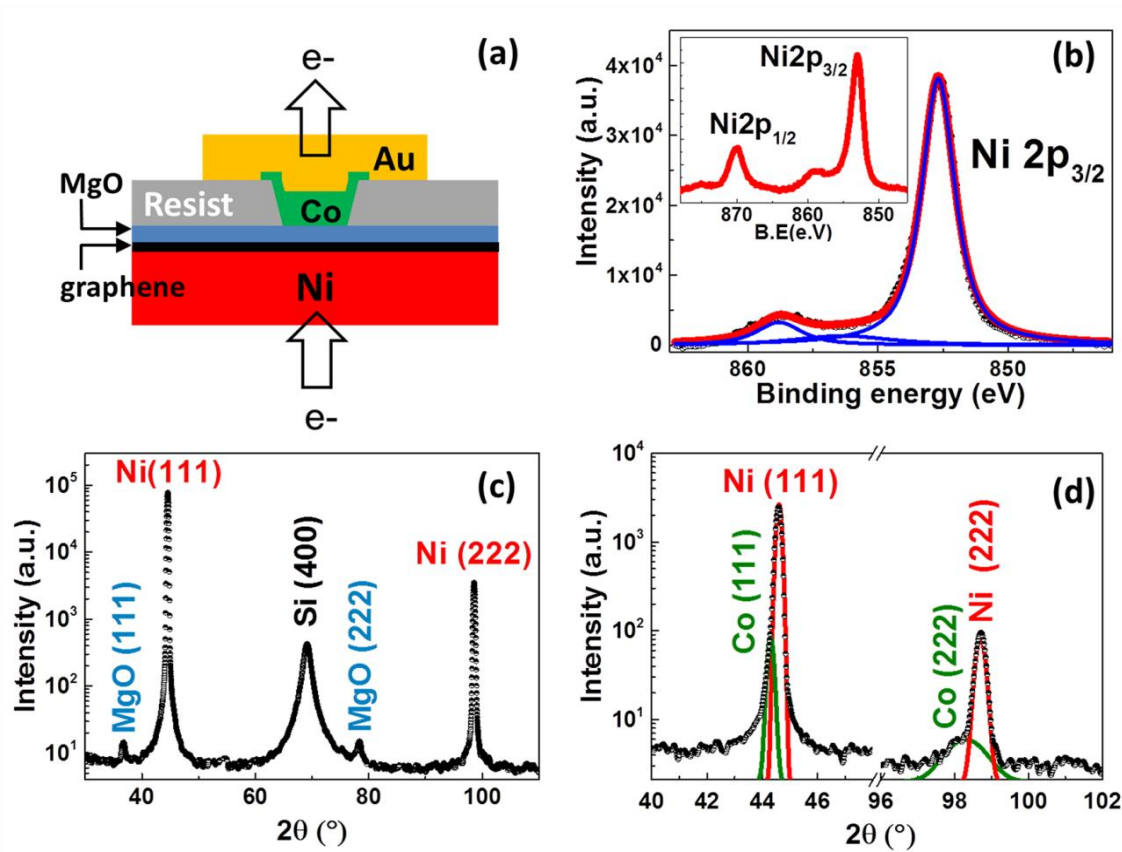

**Figure 1.** (a) Sketch of the vertical Ni/Graphene/MgO/Co spin-valve. (b) Wide (inset) and narrow energy range XPS spectra of a typical CVD Ni/Graphene sample : no oxidized state is detected, confirming the passivation of the nickel electrode by graphene. (c,d) θ-2θ x-ray diffraction spectra - λ = 1.5406 Å - obtained after deposition of MgO on the GPFE (c), and after completion of the Ni/Graphene/MgO/Co stack (d), demonstrating the (111) growth orientation of MgO, nickel and cobalt. The silicon (400) peak (c) arises from the substrate.

X-ray photon spectroscopy (XPS) measurements were used to confirm the absence of oxidation at the Gr|Ni(111) interface. Experiments were carried out using an Al $K\alpha$ X-ray source spectrometer. Figure 1(b) shows a typical spectrum centered on the Ni $2p_{3/2}$ peak and its satellites. The main peak at 852.74 eV is attributed to clean Ni metal [23] whereas the satellites peaks at +3.7 eV and +6.0 eV are well-known contributions corresponding to surface and bulk plasmons, respectively [24]. In the presence of nickel oxide (NiO), hydroxide (Ni(OH)$_2$) and oxyhydroxide (γ-NiOOH), clear XPS lines should appear at 854.7 eV, 855.3 eV and 855.8 eV respectively [23, 24, 25], and the energy difference between the Ni $2p_{3/2}$ peak and the bulk plasmon satellite should be reduced to 5.8 eV, in the presence of NiO, and to 5.3 eV, in the presence of Ni(OH)$_2$. On the other hand, in case of carbon contamination of the Ni substrate during CVD growth, a peak corresponding to the Ni-C binding energy should appear at 853 eV [26]. Our XPS spectra reveal none of these [see Fig. 1(b)] and thus confirm

the good chemical quality of the commercial GPFE substrate used. We showed previously that single crystal (111)MgO tunnel barriers can be grown on top of epitaxial graphene on (0001) SiC [17]. In the present work, we extend the applicability of this result and report the epitaxial growth of (111) oriented MgO(3 nm)/Co(50 nm) stacks on CVD Gr/(111)Ni substrate. The (111) orientation of the whole stack is clearly evidenced from the θ-2θ X-ray diffraction spectra, where only (111) and (222) peaks of cubic MgO, fcc Ni and fcc Co appear [Fig.1(c,d)].

Low temperature magneto-transport measurements were carried at the base temperature (1.5 K) of a He-flow cryostat inside a superconducting magnet, using both a high precision dc sourcemeter and a lock-in ac setup. In Fig. 2, we show the voltage dependence of the differential conductance $G(V) = dI(V)/dV$ measured on typical devices having very different junction areas of 1, 100 and 1000 $\mu m^2$, yet showing similar behaviors. The observed non-ohmic behavior is consistent with the presence of a tunnel barrier and the large resistance-area product, in the range of 10-100 MΩ.$\mu m^2$ at 200 mV, confirms the low layer roughness and the absence of pinholes through the barrier [5, 18]. We note that the G(V) characteristics of the junctions are not perfectly symmetrical, as expected from the asymmetrical composition of the stack. The dip in the G(V) curves at low bias is characteristic of electron tunneling into graphene. It corresponds to a quenching of the transmitted current due to a momentum mismatch [11, 27] : Current through graphene is expected to be carried by electrons having non-zero in-plane momentum but the probability of tunneling through a thick barrier is exponentially suppressed for such electrons which experience a larger tunneling distance. In the case of decoupled single layer graphene studied by scanning tunneling microscopy, the dip takes the form of a well-defined gap, the width of which is set by the energy necessary to open a phonon-mediated inelastic tunneling channel through the *K* point, that is ~ 67 meV (see e.g. Ref. [27]). As will be discussed later in the paper, the dip half-width is here lower - in the order of 40 meV - primarily because of the hybridization of graphene with the Ni substrate.

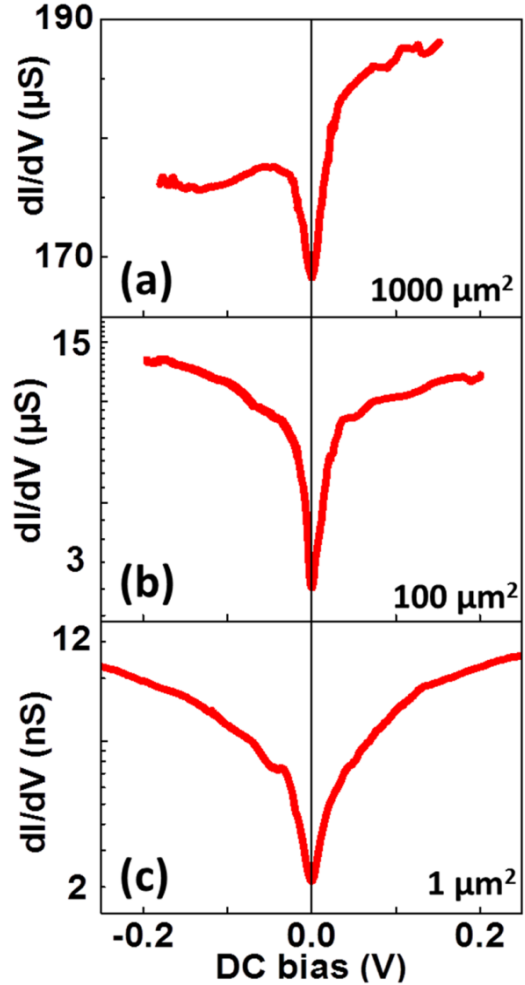

**Figure 2. Voltage dependence of the differential conductance for three vertical Ni/Graphene/MgO/Co spin-valves with junction area of 1000 μm$^2$(a), 100 μm$^2$(b) and 1 μm$^2$(c), at 1.5 K.**

Figure 3 presents the voltage dependence of the tunnel magnetoresistance ratio defined as $(R_{AP} - R_P)/R_{AP}$, with $R_P$ and $R_{AP}$ being the resistance in the parallel and antiparallel magnetic configurations respectively, for the same three junctions. These detailed TMR(V) curves showing several sign reversals were determined from I(V) curves recorded in the two magnetic states. Their reproducible character was verified by combining different couples of I(V) data sets and their accuracy was confirmed by a number of resistance versus magnetic field loops taken at different voltage values (insets in Fig. 3). The observed magnitude of TMR is similar for all samples, and the overall bias dependence behavior is qualitatively independent of the size of the device. Several regimes, named A, B, and C, can be distinguished from the bias dependence [Fig. 3]. First, at low bias - *regime A*-, a positive TMR is observed. Then, on increasing the voltage, both positively and negatively, the TMR decreases, changes sign and reaches a maximum negative value - *regime B* -. Finally, on increasing the voltage further, TMR increases again - *regime C* -, leading in some cases (mostly for negative bias) to a second sign reversal. Our data thus reveal a behavior significantly more complex than previously reported. In particular, they indicate that the sign of the spin polarization at the GPFE interface is voltage dependent.

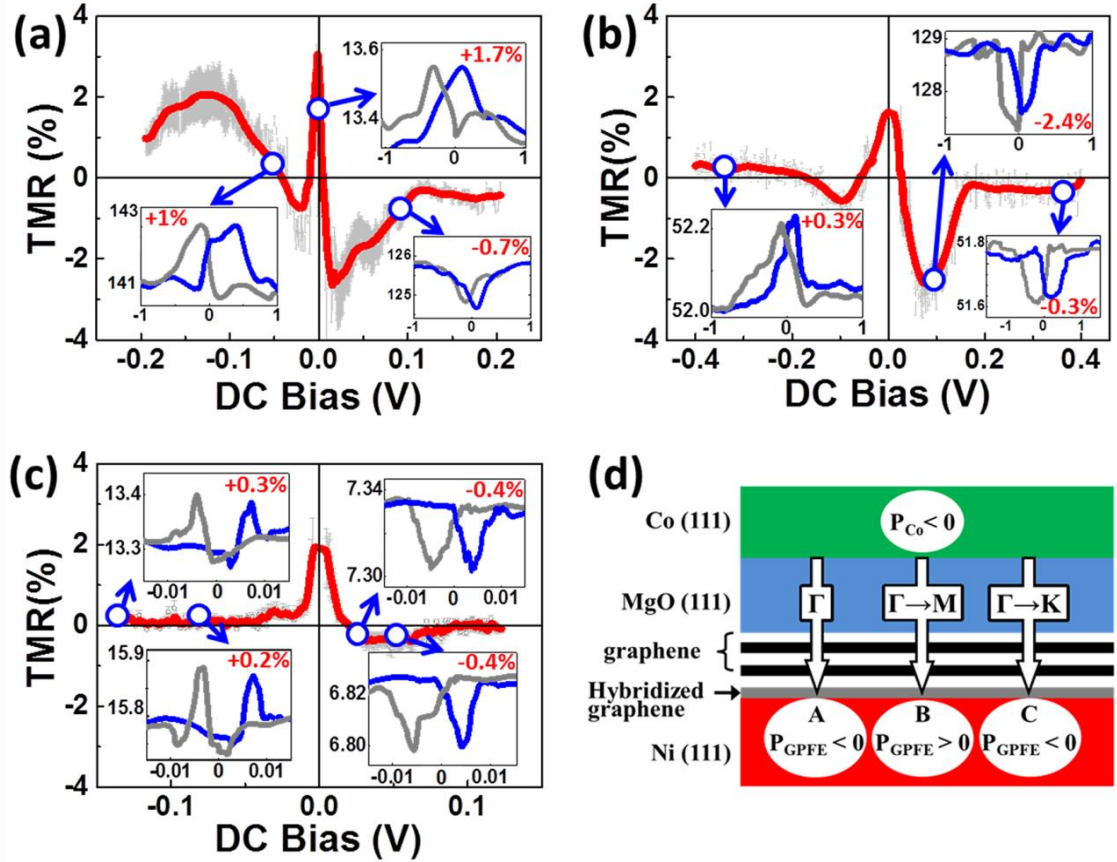

**Figure 3.** (a-c) Raw (grey) and smoothed (red) bias dependence of the tunneling magnetoresistance ratio determined from I(V) curves measured in the parallel and antiparallel magnetic configurations, for three vertical Ni/Graphene/MgO/Co spin-valves with junction area of 1000 µm$^2$ (c), 100 µm$^2$ (a) and 1 µm$^2$ (b). The insets show magnetoresistance loops (resistance in kΩ (a), MΩ (b), and kΩ (c), magnetic field in tesla) measured at different bias voltages, confirming the sign reversal of TMR. (d) Sketch of the three conduction channels that contribute to transport in vertical Ni/Graphene/MgO/Co spin-valves with their respective tunneling spin polarization : Direct tunneling through the $\Gamma$ point (A), inelastic tunneling at the *M* point (B), and inelastic tunneling at the *K* point (C).

Two phenomena may be at the origin of an inversion of the sign of the TMR: i) the resonant tunneling of electrons via localized defects in the tunnel barrier [28] and ii) the bias dependence of the tunneling spin polarization of at least one of the contacts [12, 29]. The resonant tunneling model can be reasonably discarded. Indeed it cannot account for multiple changes of sign of the TMR. Moreover the strong resemblance of the TMR(V) curves and the rather symmetrical character of its central feature would imply that defect states of similar nature be systematically present near the center of the tunnel barriers of the three distinct devices, which is highly improbable. We thus explore the likelihood of another scenario relying on the particular spin-polarized band structure of the GPFE.

Inversions of TMR with varying bias voltage have already been reported for epitaxial spin-valves [12, 13, 14]. They most often occur due to the voltage-induced opening/closing of conduction channels associated with spin-polarized electronic bands or surface states close to the Fermi level. The electronic band structure of epitaxial (111) Co is known to have only a minority spin band present at the Fermi level [7, 30, 31, 32]. This implies a negative tunneling spin polarization at the Co(111)/MgO interface ($P_{Co} < 0$). Since it is not expected to change upon varying the voltage, the changes of sign of the TMR that we observe necessarily reflect some modifications of the effective spin polarization at the surface of the GPFE ($P_{GPFE}$).

Our results can be explained in the light of recent theoretical results [7] which show that hybridization of graphene to Ni(111) produces significant modifications in its band structure : Near the Fermi level, a gap opens up at the $K$ point for majority electrons and spin polarized electronic states appear at the $\Gamma$ and $M$ points [7]. As we will discuss now, these states provide new conduction channels, which have different effective spin polarization and become active only beyond specific threshold voltages. This leads to the existence of three bias voltage regimes.

a) At low bias (*regime A*), owing to the absence of inelastic processes and the presence of a thick MgO barrier which promotes tunneling of electrons with perpendicular-to-plane momentum, conduction occurs predominantly at the $\Gamma$ point. For graphene hybridized to Ni, unlike in pristine graphene, a small density of states indeed exists at this point of the Brillouin zone [7]. It only consists of minority spin states and thus corresponds to a negative tunneling spin polarization ($P_{GPFE} < 0$). Recalling that the tunnel magnetoresistance ratio is also given by $(2\ P_{GPFE}\ P_{Co})/(1 + P_{GPFE}\ P_{Co})$, negative $P_{GPFE}$ and $P_{Co}$ yield a positive TMR, as is observed experimentally [Fig. 3].

b) Following the same line of thought, the decrease of TMR in *regime B*, which in most cases leads to a TMR sign reversal, is necessarily related to the opening of a second conduction channel with $P_{GPFE} > 0$, involving majority spin states near the Fermi level. In the Gr/Ni(111) system, such states are solely available at the $M$ point [7] and can only be reached through an inelastic process allowing electrons to circumvent the $k$-filtering effect of the thick MgO barrier. Besides, the energy of the out-of-plane acoustic phonon mode reaches about 40 meV at the *M* point [33], which is consistent with the width of the observed dip in the G(V) data curves [Fig. 2]. Although it is not clear what the sign of $P_{GPFE}$ is there, we thus propose that regime B corresponds to the activation of a phonon-mediated conduction channel through the *M* point.

c) Finally, in *regime C*, the TMR ratio tends towards positive values again. We attribute this to the activation of the well-known inelastic tunneling mechanism to *K* point states, also mediated by an out-of-plane acoustic phonon [11] [Fig. 3(d)]. Since only minority spin electron states ($P_{GPFE} < 0$) are available at the *K* point of hybridized graphene [7], this additional *K* channel provides a positive contribution to TMR. The latter may eventually dominate at large bias voltage, explaining the second TMR sign reversal we observe.

It is noteworthy that although negative TMR originating from spin filtering in GPFE based structures has been reported previously [11, 35, 37], the bias dependent tuning of sign of TMR presented here has never been demonstrated before. In previous studies using GPFE by Dlubak et al. [11], a well-defined conduction gap with half-width close to 62 mV was seen in the G(V) curve. In spite of this, some other features within the gap region are evident, which indicate other possible inelastic tunneling mechanisms occurring below 62 mV. Interestingly, also, the data are noticeably different when compared to recent reports by the same group on nominally similar GPFE-based devices [35]. It thus turns out that, while a low-bias dip/gap is systematically present, its precise shape and width may vary from sample to sample. This can originate from variations in the degree of hybridization of graphene with Ni, which has a direct influence on the energy of the phonon modes [33, 36]. Fluctuations in the degree of hybridization could also modify the relative contributions of the three conduction channels (*Γ*, *K,* and *M*) to the overall spin polarized transport and be responsible for the fact that regimes A, B, and C do not seem to span exactly the same voltage ranges in the three samples considered in the present study.

In conclusion, our study of the voltage-dependent magneto-transport properties of GPFE/MgO/Co vertical spin-valves with thick MgO tunnel barriers reveals the complexity of the spin-filtering effects at the (111) Ni|Gr hybrid interface. The good crystalline quality of the studied devices allows for the identification of three distinct conduction channels and the determination of the sign of the tunneling spin polarization at the Ni|Gr interface, in each of them. Based on recent theoretical predictions of spin polarized band structure of the

(111)Ni|Gr hybrid interface, we propose that the three channels correspond to electron transport through three distinct regions of the Brillouin zone, namely the $\Gamma$, $M$, and $K$ points. Conduction through both the $\Gamma$ point and the $M$ point appears specifically as a result of the hybridization of graphene with Ni but while conduction through the $\Gamma$ point occurs by direct tunneling, conduction through the $M$ point is mediated by phonons, as conduction through the $K$ point. The tunnel spin polarization of the GPFE changes from negative in the $\Gamma$ and $K$ channels to positive in the $M$ channel, the latter dominating conduction at intermediate voltage values. This gives rise to the observed multiple sign inversions of TMR upon varying the voltage. More generally, our work demonstrates that tailoring the tunneling spin polarization of well-known transition metals through hybridization with two-dimensional overlayers provides opportunities for realizing new spin injectors with unique bias-dependent properties.

Acknowledgements : We thank Fabien Chevrier and STnano cleanroom staff for technical support, Alain Carvalho and Matthieu Bailleul for technical discussion during development of the ebeam process, Stephane Berciaud for fruitful discussions on electron-phonon coupling mechanisms in graphene, Guy Schmerber for XRD measurements, and Thierry Dintzer of ICPEES for XPS measurements. Financial supports from CNRS (Nano 2012: G3N) and from ANR (FUNGRAPH 11-IS10-003 02, Labex NIE ANR-11-LABX-0058_NIE) are also gratefully acknowledged.